\documentclass[aps,pra,showpacs,twocolumn,amsmath,amssymb]{revtex4-1}
\usepackage{graphicx}
\usepackage{amsmath}
\usepackage{color}
\usepackage{dcolumn}
\usepackage{bm}
\usepackage{booktabs}
\usepackage{subfigure}

\newcommand{\SM}{Suppl.\,Mat.\,}
\newcommand{\ourtitle}{Quantum-State Controlled Penning Ionization Reactions between Ultracold Alkali and Metastable Helium Atoms}

\begin{document}

\title{\ourtitle}

\author{A.\,S.\,Flores}
\author{W.\,Vassen}
\author{S.\,Knoop}
\affiliation{LaserLaB, Department of Physics and Astronomy, Vrije Universiteit, De Boelelaan 1081, 1081 HV Amsterdam, The Netherlands}

\date{\today}

\begin{abstract}
In an ultracold, optically trapped mixture of $^{87}$Rb and metastable triplet $^4$He atoms we have studied trap loss for different spin-state combinations, for which interspecies Penning ionization is the main two-body loss process. We observe long trapping lifetimes for the purely quartet spin-state combination, indicating strong suppression of Penning ionization loss by at least two orders of magnitude. For the other spin-mixtures we observe short lifetimes that depend linearly on the doublet character of the entrance channel. We compare the extracted loss rate coefficient with recent predictions of multichannel quantum-defect theory for reactive collisions involving a strong exothermic loss channel and find near-universal loss for doublet scattering. Our work demonstrates control of Penning ionization reactive collisions by internal atomic state preparation. 
\end{abstract}

\maketitle
Ultracold inelastic and reactive collisions are important processes in atomic and molecular samples \cite{julienne1989cou,quemener2012umu}, determining their trapping lifetimes and the success of evaporative and sympathetic cooling. Conversely, measurements of these lifetimes reveal the rate coefficients of the dominant inelastic or reactive collision processes, opening the fields of ultracold few-body physics \cite{kraemer2006efe,knoop2009ooa} and ultracold chemistry \cite{ospelkaus2010qsc,knoop2010mce}. The ultracold regime offers exquisite control over the initial internal and external quantum states, and the possibility to experimentally control collision properties or even steer chemical reactions with external fields \cite{krems2008ccc}.

Understanding of inelastic and reactive collisions is in general very difficult due to the many degrees of freedom involved. This has motivated recent work based on multichannel quantum-defect theory (MQDT) \cite{idziaszek2010urc,gao2010umf,jachymski2013qto}, in which analytic expressions of collision rates were derived in the case of a strong exothermic reactive channel. In particular, if the probability of an inelastic or reactive process in the short-range part of the collision is 100\%, i.\,e.\,if $P^{\rm re}=1$, theory predicts universal rate constants that only depend on the reduced mass of the collision partners and the leading long-range coefficient \cite{idziaszek2010urc,gao2010umf}, independent of the complicated short-range dynamics. If the reaction probability is less than 100\% ($P^{\rm re}<1$), still only two parameters are required to include the (non-universal) short-range physics, i.\,e.\,the scattering length $a$ and $P^{\rm re}$ \cite{jachymski2013qto}. These analytical models have been applied to atom-exchange reactions between ground state KRb molecules below 1~$\mu$K \cite{ospelkaus2010qsc,idziaszek2010urc}, and Penning ionization reactions between argon and helium atoms in the metastable triplet 2~$^3$S$_1$ state (He$^*$) in merged-beam experiments from 10 mK up to 30~K \cite{henson2012oor,jachymski2013qto}.

\begin{figure}
\includegraphics[width=8.5cm]{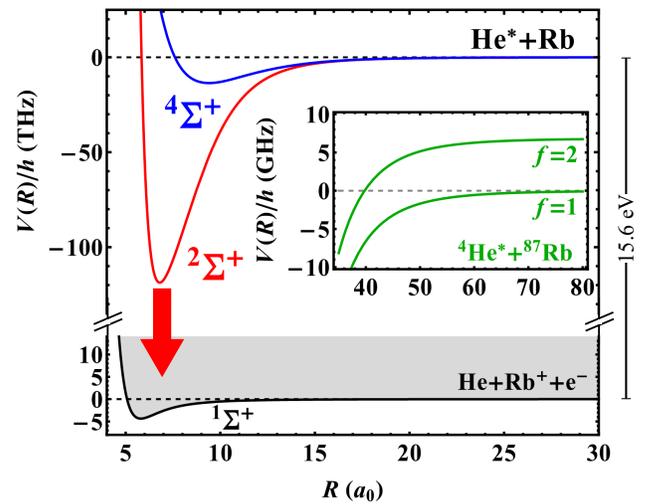}
\caption{(Color online) Potential energy curves of the $^2\Sigma^+$ \cite{ruf1987tio} and $^4\Sigma^+$ \cite{knoop2014umo} states of the He*Rb molecule that correlate with the He$^*$+Rb atomic asymptote, and the $^1\Sigma^+$ \cite{hickling2004som} state of the HeRb$^+$ molecule that correlates with the He+Rb$^+$+e$^-$ asymptote, which lies 15.6~eV lower and forms the Penning ionization continuum (where the internuclear distance $R$ is given in Bohr radii, $a_0=0.05292$~nm). The inset shows the long-range adiabatic potentials of $^4$He*$^{87}$Rb near its dissociation threshold, including the hyperfine splitting of $^{87}$Rb.  
\label{molpot}}
\end{figure}

In this Rapid Communication we study ultracold Penning ionizing collisions between He$^*$ atoms (internal energy 19.8~eV) and alkali atoms A in their electronic ground state: 
\begin{equation}
{\rm He}(2 ^3{\rm S}_1)+{\rm A}(^2{\rm S}_{1/2}) \rightarrow {\rm He}(1 ^1{\rm S}_0)+{\rm A}^+(^1{\rm S}_0)+{\rm e}^-\label{PI},
\end{equation}
which are described by two interaction potentials, doublet $^2\Sigma^+$ and quartet $^4\Sigma^+$, and a strongly exothermic Penning ionization (PI) reaction channel (see Fig.~\ref{molpot} for the specific case of He$^*$+Rb). The description of the PI loss rate in terms of MQDT would require at least four parameters, namely the scattering lengths and the reaction probabilities of both doublet and quartet potentials. However, PI from the $^4\Sigma^+$ potential is spin-forbidden, because the total electron spin in the PI channel is only 1/2 (see Eq.~\ref{PI}). Thus PI proceeds predominately via the doublet $^2\Sigma^+$ potential. Therefore one expects the PI loss rate to be determined by the $^2\Sigma^+$ potential only, however, including an additional factor that takes into account the doublet character of the particular entrance channel. This makes the PI loss rate experimentally controllable by internal atomic state preparation and magnetic field.

\begin{figure}
\includegraphics[width=8.5cm]{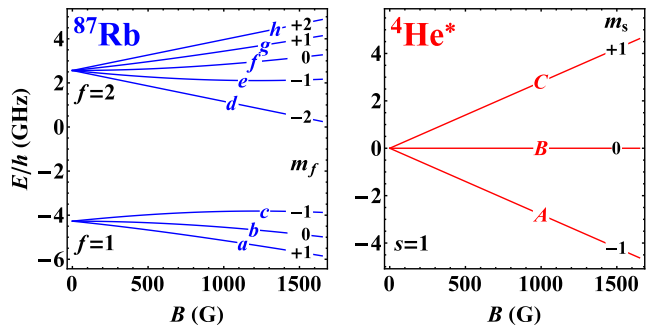}
\caption{(Color online) Magnetic field dependence of the atomic ground-state energies of $^{87}$Rb and $^4$He$^*$, indicating the labeling of the different internal states as used throughout the paper.
\label{Zeeman}}
\end{figure}

We have realized an ultracold mixture of $^4$He$^*$ and $^{87}$Rb in an optical dipole trap (ODT), and performed lifetime measurements for different spin-state combinations (the labeling of the atomic spin-states is shown in Fig.~\ref{Zeeman}, ($a$-$h$) for $^{87}$Rb and ($A$-$C$) for $^4$He$^*$). ODTs provide spin-independent confinement, applicable to both low- and high-field seeking spin-states, which allows direct comparison between trap losses of different spin-mixtures. Previous experimental studies of He$^*$+alkali collisions have been performed at thermal energies in stationary afterglow and merged-beam experiments (see e.\,g.\,\cite{johnson1978pio,ruf1987tio}). Simultaneous laser cooling and trapping of $^4$He$^*$ and $^{87}$Rb was first demonstrated by the Truscott group \cite{byron2010tli}. Magnetic trapping of the $h+C$ spin-state combination, which is purely quartet, provided upper limits of the PI rate on the order of $10^{-12}$~cm$^3$s$^{-1}$ for pure quartet scattering \cite{byron2010sop,knoop2014umo}, and revealed a small quartet scattering length $a_Q$ \cite{knoop2014umo}, in agreement with \emph{ab initio} calculations of the quartet $^4\Sigma^+$ potential \cite{knoop2014umo,kedziera2015aii}. In contrast, the knowledge on the doublet $^2\Sigma^+$ potential is limited \cite{ruf1987tio}, and the doublet scattering length $a_D$ is unknown. 

Starting point of our measurements is an ultracold mixture of $3\times10^4$ $^4$He$^*$ and $9\times10^4$ $^{87}$Rb atoms in a single-beam ODT at a temperature of 22 and 15~$\mu$K, respectively. The main parts of our experimental setup are described earlier: dual-species magneto-optical trap and transfer to quadrupole magnetic trap (QMT) \cite{knoop2014umo} and production of $^{87}$Rb \cite{mishra2015epo} and $^4$He$^*$ \cite{flores2015smf} Bose-Einstein condensates using a single-beam ODT. Here we apply simultaneous RF- and MW-forced evaporative cooling in the QMT for He$^*$ and Rb, respectively, before transfer to the single-beam ODT, which has a waist of 40~$\mu$m and a wavelength of 1557~nm. We use a fixed ODT power of 3.8~W, corresponding to an effective trap depth of 200~$\mu$K and 140~$\mu$K for He$^*$ and Rb, respectively \footnote{The polarizability of He$^*$ is 1.4 times larger than Rb at a wavelength of 1557~nm \cite{safronova2006fdp,notermans2014mwf}, while at our ODT power gravity gives a 5\% reduction on the Rb trap depth (which for He$^*$ is well below 1\%). The interspecies thermalization rate is small (0.01~s$^{-1}$) due to the small interspecies scattering length \cite{knoop2014umo} and the large mass ratio, and only intraspecies thermalization takes place (for which the rate is about 1~s$^{-1}$ for both species).}. 

Throughout the preparation stages in the QMT and ODT we use the stable $h+C$ spin-state combination \cite{knoop2014umo}. To prepare other spin-mixtures we transfer $^{87}$Rb from $h$ to $a$ and/or $^4$He$^*$ from $C$ to $A$, by single adiabatic MW and RF frequency sweeps, respectively, at a bias magnetic field of 2.5~G. While our RF transfer has a 100\% efficiency, our MW transfer is only 50\% due to limited MW power. We remove the non-transferred Rb atoms in state $h$ with resonant light immediately after the MW sweep. After a variable hold time we switch off the ODT and simultaneously measure the number of atoms by using standard absorption imaging for Rb and microchannel plate (MCP) detection for He$^*$ \cite{vassen2012cat}. 

We obtain the interspecies Penning ionization loss rate coefficients by measuring the time-evolution of the number of He$^*$ atoms, and fit the solution of two coupled equations:
\begin{equation}\label{PIloss2}
\dot{N}_{i}=-\Gamma_{i} N_{i}-L_2 \int n_{i}(\vec{r})n_{j}(\vec{r}) {\rm d}\vec{r}
\end{equation}
where $(i,j)$ is (He$^*$, Rb), $L_2$ is the total interspecies two-body loss rate coefficient, $N_i$ and $n_i(\vec{r})$ are the atom number and density profile for species $i$. Intraspecies two- and three-body loss processes \cite{soding1999tbd,marte2002fri,borbely2012mfd} can be fully neglected for the chosen spin-states under our conditions. We only fit the time-evolution of He$^*$, using the measured initial Rb atom number, because of the higher sensitivity being the minority species and the better signal-to-noise of the He$^*$ MCP detection compared to the absorption imaging of Rb. The density profiles are calculated numerically, using measured temperatures $T_i$, via $n_{i}(\vec{r})=n_i^0\exp\left[-U_i(\vec{r})/k_B T_i\right]$, where $U_i(\vec{r})$ is the trapping potential, including gravity, and $n_i^0=N_i/\int{\exp\left[-U_i(\vec{r})/k_B T_i\right]{\rm d}\vec{r}}$ is the peak density. In our case the vertical confinement is strong enough such that the reduction of the overlap between the two clouds due to the differential gravitional sag is negligible. Single-species lifetimes are measured to determine the one-body loss rates $\Gamma_i$.  

\begin{figure}
\includegraphics[width=8.5cm]{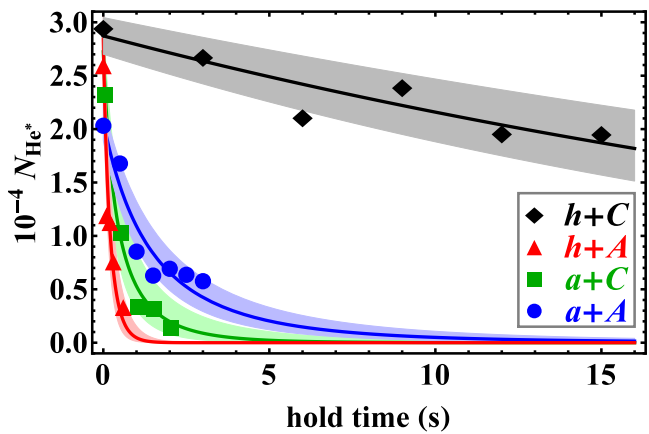}
\caption{(Color online) Time-evolution of the number of He$^*$ atoms for different spin-mixtures, at a bias magnetic field of 2.5~G. The displayed data represent an average over several experimental runs (3-6). The solid lines are fits of Eq.~\ref{PIloss2} to all the data. The colored bands around the lines indicate the standard error of the fit. The initial Rb atom numbers are: $h+A$ and $h+C$: $N_{\rm Rb}=(8.6\pm0.7)\times 10^4$; $a+A$: $N_{\rm Rb}=(4.5\pm0.8)\times 10^4$; $a+C$: $N_{\rm Rb}=(3.9\pm0.6)\times 10^4$.
\label{lifetime}}
\end{figure}

In Fig.~\ref{lifetime} we present our lifetime measurements (at the bias magnetic field of 2.5~G), showing the time-evolution of the He$^*$ atom number of the different spin-mixtures. We observe a long trapping lifetime for the purely quartet $h+C$ spin-combination, which we cannot distinguish from the single-species lifetimes. This means that the trapping lifetime is fully dominated by one-body loss and we can only obtain an upper limit of the two-body loss rate, namely $1.3\times10^{-12}$~cm$^3$s$^{-1}$. Together with our knowledge of $a_Q$ \cite{knoop2014umo}, we obtain a constraint on the reaction probability for the quartet $^4\Sigma^+$ potential of $P^{\rm re}<0.01$, using Eq.~\ref{nurc} below. We expect the actual quartet PI loss rate to be on the order of $10^{-14}$~cm$^3$s$^{-1}$, on basis of the suppression of PI in homonuclear He$^*$ collisions \cite{vassen2012cat} and the similar $s$ character of the valence electron of He$^*$ and alkali atoms. In the following we simply neglect the quartet contribution to the PI loss. 

\begin{figure}
\includegraphics[width=8.5cm]{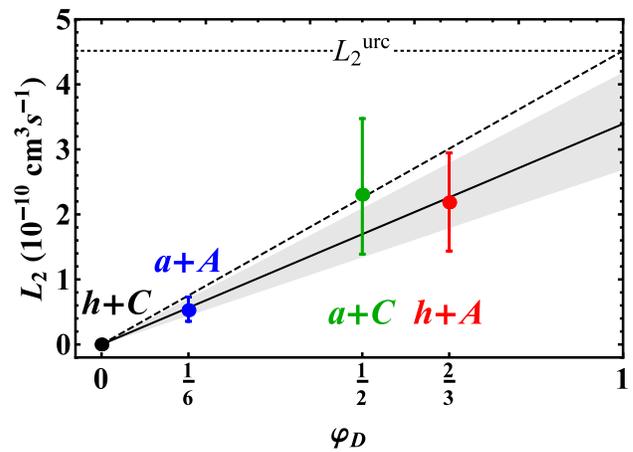}
\caption{(Color online) Compilation of measured two-body loss rates $L_2$, sorted by the corresponding doublet character $\varphi_D$, and comparison with the universal loss rate $L_2^{\rm urc}$ (dotted line) via $L_2=\varphi_D L_2^{\rm urc}$ (dashed line). The solid line is a linear fit through the data points (and its colored band indicates the standard error of the fit).
\label{lossrates}}
\end{figure}

For the $h+A$, $a+C$ and $a+A$ spin-mixtures we observe orders of magnitude faster losses, and especially for the $h+A$ and $a+C$ spin-mixtures the He$^*$ sample is depleted within a few seconds. Here one should note the approximately factor of two difference in initial Rb atom numbers between the $h+A$ mixture and the $a+A$ and $a+C$ mixtures, mainly due to the MW transfer efficiency. We obtain the two-body loss rate coefficients $L_2$ by fitting Eq.~\ref{PIloss2} to the data, and the results are shown in Fig.~\ref{lossrates}. The error bars contain the fit error and uncertainty in the initial Rb atom number, as well as the uncertainty in the temperatures, which are required to calculate the density profile. 

First we analyze our data in terms of the doublet character $\varphi_D$, which represents the amount of doublet scattering. Provided that PI is the dominant loss process, we expect the loss rate to scale linearly with $\varphi_D$,
\begin{equation}\label{PD}
L_2=\varphi_D L_2^{\rm PI},
\end{equation}
where $L_2^{\rm PI}$ is the loss rate due to PI for pure doublet scattering. $\varphi_D$ is obtained by expanding the long-range atomic product states on to the short-range doublet molecular state (see \SM~\cite{supp}). In the limit of low magnetic fields, i.\,e.\,$B<<E_{\rm HFS}/4\mu_B$ (where $E_{\rm HFS}$ is the alkali hyperfine splitting and $\mu_B$ is the Bohr magneton), and the case of an alkali atom with nuclear spin of 3/2, $\varphi_D$ takes values of $q/6$, where $q$ is an integer number between 0 and 4. The value of $\varphi_D$ for the different spin-state combinations is indicated in Fig.~\ref{lossrates}. Fitting Eq.~\ref{PD} to our data gives $L_2^{\rm PI}=3.4^{+0.8}_{-0.7}\times 10^{-10}$~cm$^3$s$^{-1}$ for Penning ionization loss via the doublet potential. 

Second, we compare the obtained value of $L_2^{\rm PI}$ with analytic expressions from MQDT. For the universal case, $P^{\rm re}=1$, the zero-temperature limit universal loss rate is given by \cite{idziaszek2010urc,gao2010umf}:
\begin{equation}\label{urc}
L_2^{\rm urc}=2\frac{h}{\mu}\bar{a}
\end{equation}
where $\bar{a}=0.478\ldots\left(2\mu C_6/\hbar^2\right)^{1/4}$ is the so-called mean scattering length that solely depends on the reduced mass $\mu$ and the leading long-range van der Waals coefficient $C_6$. For He$^*$+Rb $C_6=3858$~a.u. \cite{knoop2014umo}, resulting for $^4$He$^*$+$^{87}$Rb in $\bar{a}=41a_0$ and $L_2^{\rm urc}=4.5\times 10^{-10}$~cm$^3$s$^{-1}$. In Fig.~\ref{lossrates} $L_2^{\rm urc}$ and $\varphi_D L_2^{\rm urc}$ are shown as the dotted and dashed lines, respectively. Our extracted value of $L_2^{\rm PI}$ lies slightly below $L_2^{\rm urc}$. However, taking into account a small finite temperature correction of 8\% reduces the universal loss rate to $4.2\times 10^{-10}$~cm$^3$s$^{-1}$ (see \SM~\cite{supp}), which is consistent with our $L_2^{\rm PI}$ and the reaction probability for PI in the doublet potential might be 100\%.

\begin{figure}
\includegraphics[width=8.5cm]{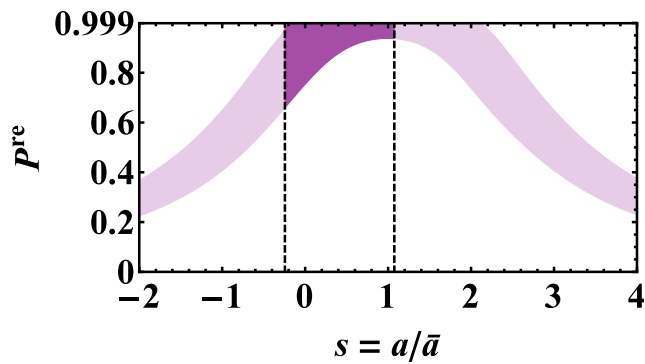}
\caption{(Color online) Comparison between the measured PI loss rate and the MQDT prediction for non-universal loss (Eq.~\ref{nurc} including finite temperature corrections), where the purple colored band represents the combinations of $P^{\rm re}$ and $s$ that matches $L_2^{\rm PI}$. The vertical dashed lines give the bounds on the scattering length from our analysis of hyperfine changing collisions, which constraints the possible combinations to the darker part of the purple colored band.
\label{MQDT_Pre_a}}
\end{figure}

Still, $L_2^{\rm PI}$ could also correspond to the non-universal case, $P^{\rm re}<1$, where the zero-temperature limit loss rate is given by \cite{jachymski2013qto}: 
\begin{equation}\label{nurc}
L_2^{\rm nurc}=L_2^{\rm urc}y\frac{1+\left(s-1\right)^2}{1+y^2\left(s-1\right)^2},
\end{equation}
where $y$ is related to $P^{\rm re}$ via $P^{\rm re}=4y/\left(1+y\right)^2$, and $s=a/\bar{a}$ is the rescaled scattering length, where $a$ is the scattering length (here the doublet scattering length $a_D$). The combinations of $P^{\rm re}$ and $s$ that matches $L_2^{\rm PI}$ are shown in Fig.~\ref{MQDT_Pre_a} as the purple colored band (for which we have also included finite temperature corrections). A typical scattering length of $a\approx\bar{a}$, i.\,e.\,$s\approx 1$, would mean a high reaction probability of $P^{\rm re}\gtrsim 0.94$. While for $P^{\rm re}=1$ sensitivity to the scattering length is lost (and $L_2^{\rm nurc}=L_2^{\rm urc}$), a tiny reduction of less than 0.001 already results in a constraint on the possible scattering length range of about $0\lesssim s \lesssim 2$. A match with a small reaction probability would require the less likely cases of either a large positive or large negative scattering length.

In principle, $P^{\rm re}$ can be calculated from the complex potential $U(r)=V(r)-(i/2)\Gamma(r)$ \cite{miller1970top}, where $\Gamma(r)$ represent the ionization width. For He$^*$+alkali collisions \emph{ab initio} calculations on the doublet potential $V(r)$ and corresponding $\Gamma(r)$ are available for Li, Na and K \cite{cohen1985tco,*scheibner1987roa,*merz1990eat,*movre2000tio}, from which one can estimate $P^{\rm re}\approx 0.6-0.7$. For He$^*$+Rb information on the doublet potential $V(r)$ is limited \cite{ruf1987tio}, while $\Gamma(r)$ is completely lacking. Stationary afterglow experiments at thermal energies \cite{johnson1978pio} gave an anomalously large loss rate for Rb, compared to Na, K and Cs, and from this loss rate one estimates $P^{\rm re}\approx 1$ using the classical Gorin model \cite{jachymski2014qdm} and assuming a statistical weight of $1/3$ for doublet scattering. The presence of autoionizing Rb states close to the excitation energy of He$^*$ was suggested as a possible explanation of this anomaly \cite{johnson1978pio}, however, no fingerprint of these additional ionization channels was found in the electron emission spectrum \cite{ruf1987tio}. \emph{Ab initio} calculations of $V(r)$ and $\Gamma(r)$ are required to resolve this issue, however, it is probably safe to assume that $P^{\rm re}>0.5$.

While for $a+A$ and $a+C$ PI is the only exothermic, spin-allowed two-body loss process, for $h+A$ hyperfine changing collisions (HCC) provide an additional two-body loss channel. However, comparing the loss rates for $a+C$ and $h+A$ suggests that the HCC contribution is small. By determining $L_2^{\rm PI}$ on basis of $a+A$ and $a+C$ only, and obtaining a lower limit of the PI contribution for $h+A$ via $\varphi_D L_2^{\rm PI}$, we derive an upper limit of HCC loss rate of $1.2\times 10^{-10}$~cm$^3$s$^{-1}$. According to the theory of ultracold spin-exchange collisions, as derived for hydrogen or alkali atoms, the HCC loss rate depends on the difference between the scattering lengths of the two interaction potentials. Applying the analytical result of Ref.~\cite{stoof1988sea} directly to He$^*$+alkali collisions, with our HCC upper limit we derive that $|a_D-a_Q|<23a_0$ (see \SM~\cite{supp}). With our previously determined value $a_Q=17(4)a_0$ \cite{knoop2014umo}, this would corresponds to $-10a_0<a_D<44a_0$. This constraint significantly reduces possible values of $P^{\rm re}$, as indicated by the darker part of the purple colored band in Fig.~\ref{MQDT_Pre_a}. However, this analysis assumes that there is no influence of the PI channel on the HCC process, which may be too simplistic. 

In general the doublet character $\varphi_D$ is magnetic field dependent, and therefore also the loss rate. For instance, for the energetically lowest spin-channel $a+A$ $\varphi_D$ becomes significantly less than $1/6$ when $B\sim E_{\rm HFS}/4\mu_B$, and $\varphi_D \rightarrow 0$ as $B^{-2}$ for $B>>E_{\rm HFS}/4\mu_B$  (see \SM~\cite{supp}). While for $^{87}$Rb this behavior occurs at rather high magnetic fields ($E_{\rm HFS}/4\mu_B=1.2$~kG), due to the large $E_{\rm HFS}$, for an alkali atom with a small $E_{\rm HFS}$, like $^{41}$K, this effect takes place within an experimentally accessible range of magnetic fields ($E_{\rm HFS}/4\mu_B=45$~G). This provides interesting prospects for realizing stable ultracold mixtures in a variety of spin-state combinations, and the application of Feshbach resonances to tune the scattering length, which requires small two-body losses \cite{hutson2007fri}. 

In conclusion, we have realized an ultracold, optically trapped mixture of $^4$He$^*$ and $^{87}$Rb atoms and obtained the two-body loss rate coefficients for four different spin-mixtures. We find long trapping lifetimes for the purely quartet spin-state combination, indicating a strong suppression of Penning ionization by at least two orders of magnitude, providing good prospects of realizing dual Bose-Einstein condensates. For the other spin-mixtures we observe short lifetimes that depend on the doublet character, which suggests suppression of Penning ionization at higher magnetic fields, experimentally feasible for alkali atoms with a small hyperfine splitting. We have compared our measured loss rates with recent predictions of MQDT for reactive collisions involving a strong exothermic loss channel. We observe near-universal loss for the doublet potential, and obtain a constraint on the unknown doublet scattering length.

Ultracold collisions between He$^*$ and alkali atoms can exhibit magnetically-induced Feshbach resonances \cite{chin2010fri} due to the hyperfine coupling between the doublet $^2\Sigma^+$ and quartet $^4\Sigma^+$ potentials. In combination with PI these atomic collision systems provide a relatively simple and experimentally feasible platform to study the effect of a strong exothermic loss channel on Feshbach resonances \cite{hutson2007fri,hutson2009dri}, which may be important for evaporative and sympathetic cooling for molecules. Our analysis of the Penning ionization loss rate assumes no coupling between the doublet and quartet interaction potentials, which is corroborated by the observed linear dependence of the doublet character. However, around interspecies Feshbach resonances we expect a breakdown of this simple scaling, which opens the possibility of Feshbach spectroscopy despite strong two-body losses. More elaborate MQDT calculations \cite{idziaszek2011mqd} or numerical coupled-channel calculations using \emph{ab initio} potentials are needed to investigate the behavior of the Penning ionization loss rate around these Feshbach resonances.

We acknowledge Rob Kortekaas for excellent technical support. We thank Piotr \.Zuchowski, Hartmut Hotop, Paul Julienne, Bo Gao and Krzystof Jachymski for fruitful discussions. This work was financially supported by the Netherlands Organization for Scientific Research (NWO) via a VIDI grant (680-47-511) and the Dutch Foundation for Fundamental Research on Matter (FOM) via a Projectruimte grant (11PR2905). 

%

%%%%%%%%%% Merge with supplemental materials %%%%%%%%%%

\onecolumngrid
\clearpage
\pagebreak
\begin{center}
\textbf{\large Supplemental Materials: \\ \ourtitle}
\end{center}
%%%%%%%%%% Merge with supplemental materials %%%%%%%%%%
%%%%%%%%%% Prefix a "S" to all equations, figures, tables and reset the counter %%%%%%%%%%
\setcounter{equation}{0}
\setcounter{figure}{0}
\setcounter{table}{0}
\makeatletter
\renewcommand{\theequation}{S\arabic{equation}}
\renewcommand{\thefigure}{S\arabic{figure}}
%\renewcommand{\bibnumfmt}[1]{[S#1]}
%\renewcommand{\citenumfont}[1]{S#1}
%%%%%%%%%% Prefix a "S" to all equations, figures, tables and reset the counter %%%%%%%%%%

\section{Doublet character}

The doublet $\varphi_D$ and quartet $\varphi_Q$ character of the different spin-state combinations can be obtained by expanding the long-range atomic product states on to the short-range doublet and quartet molecular states, respectively. In the low and high magnetic field limits, $\varphi_S$ ($S=D$ or $Q$) is obtained by standard angular momentum algebra. In the following we consider the general case of a pair of atoms with electron spin $s_a$ and nuclear spin $i_a$, and their respective projections $m_{s_a}$ and $m_{i_a}$, where $a=(1,2)$. The hyperfine quantum numbers and their projections are $f_a=s_a+i_a$ and $m_{f_a}=m_{s_a}+m_{i_a}$. The molecular electron spin is $S=s_1+s_2$, and its projection $m_S=m_{s_1}+m_{s_2}$. For the case of He$^*$+alkali atoms, $s_1=1$, $s_2=1/2$, and $S=1/2$ (doublet potential) or $S=3/2$ (quartet potential).

In the high magnetic field limit (Paschen-Back regime), $B>>E_{\rm HFS}/\mu_B$ (where $E_{\rm HFS}$ is the largest hyperfine splitting of the two atoms), the atomic states are labeled $|m_{s_1}, m_{i_1}; m_{s_2}, m_{i_2}\rangle$. They are projected on the molecular states via the following expansion:
\begin{equation}\label{molecularexpansion}
|S, m_{S}\rangle=\sum_{m_{s_1}=-s_1}^{+s_1}\sum_{m_{s_2}=-s_2}^{+s_2}{C_{s_1, m_{s_1}; s_2, m_{s_2}}^{S, m_S} |s_1, m_{s_1}; s_2, m_{s_2}\rangle},
\end{equation}
with Clebsch-Gordan coefficients
\begin{equation}\label{CG}
C_{s_1, m_{s_1}; s_2, m_{s_2}}^{S, m_S}=(-1)^{-s_2+s_1-m_S}\sqrt{2S+1}\left(\begin{array}{ccc}
s_1 & s_2 & S\\
m_{s_1} & m_{s_2} & -m_{S}
\end{array} \right),
\end{equation}
and the doublet and quartet character is simply given by
\begin{equation}
\varphi_S^{\rm PB}\left(m_{s_1},m_{s_2}\right)=\sum_{m_S=-S}^{+S}{\left|C_{s_1, m_{s_1}; s_2, m_{s_2}}^{S, m_S}\right|^2},
\end{equation}
independent of the nuclear spins. Therefore this result is the same for each He$^*$+alkali combination, and $\varphi_D=0$ in case $|m_{s_1}+m_{s_1}|=3/2$, and $\varphi_D=1/3$ or 2/3 otherwise.

In the low magnetic field limit (Zeeman regime), $B<<E_{\rm HFS}/\mu_B$ (where $E_{\rm HFS}$ is the smallest hyperfine splitting of the two atoms), the atomic states are labeled with the hyperfine quantum numbers, $|f_1, m_{f_1}; f_2, m_{f_2}\rangle$. For each atom the hyperfine states are given by:
\begin{equation}
|f, m_{f}\rangle=\sum_{m_{s}=-s}^{+s}\sum_{m_{i}=-i}^{+i}{C_{s, m_s; i, m_i}^{f, m_f} |s, m_{s}; i, m_i\rangle},
\end{equation}
involving the Clebsch-Gordan coefficients (Eq.~\ref{CG}). To obtain the characters first one has to project $|f_1, m_{f_1}; f_2, m_{f_2}\rangle$ states on to the $|s_1, m_{s_1}; s_2, m_{s_2}\rangle$ states, which are subsequently projected on the molecular states, as in Eq.~\ref{molecularexpansion}. The final result is:
\begin{equation}\label{fSZR}
\varphi_S^{\rm ZR}\left(f_1, m_{f_1}; f_2, m_{f_2} \right)=\sum_{m_{s_1}=-s_1}^{+s_1}\sum_{m_{s_2}=-s_2}^{+s_2}\sum_{m_S=-S}^{+S} 
{\left|C_{s_1, m_{s_1}; i_1, m_{i_1}}^{f_1, m_{f_1}}C_{s_2, m_{s_2}; i_2, m_{i_2}}^{f_2, m_{f_2}}C_{s_1, m_{s_1}; s_2, m_{s_2}}^{S, m_S}\right|^2}
\end{equation}
with $m_{i_a}=m_{f_a}-m_{s_a}$. Here $\varphi_S^{\rm ZR}$ depends explicitly on $i_1$ and $i_2$, and therefore on the particular He$^*$  and alkali isotopes. In all cases, for the doubly spin-stretched state combination, $|f_1^{\rm max}, \pm f_1^{\rm max}; f_2^{\rm max}, \pm f_2^{\rm max}\rangle$, $\varphi_D=0$.

In case of $^4$He$^*$, which has no nuclear spin (i.\,e.\,$i_1=0$), only $m_{s_1}$ determines the spin-state of $^4$He$^*$ and Eq.~\ref{fSZR} is simplified to:
\begin{equation}\label{fSZR4He}
\varphi_S^{\rm ZR, ^4{\rm He}^*}\left(m_{s_1}; f_2, m_{f_2} \right)=\sum_{m_{s_2}=-s_2}^{+s_2}\sum_{m_S=-S}^{+S}{\left|C_{s_2, m_{s_2}; i_2, m_{i_2}}^{f_2, m_{f_2}}C_{s_1, m_{s_1}; s_2, m_{s_2}}^{S, m_S}\right|^2}.
\end{equation}

At intermediate magnetic fields the expansion of the $|f, m_{f}\rangle$ states in terms of $|s, m_{s}; i, m_i\rangle$ states is magnetic field dependent, and one has to numerically calculate the eigenvectors of the Hamiltonian that contains the hyperfine and Zeeman interactions. 

\begin{figure}
\includegraphics[width=8.5cm]{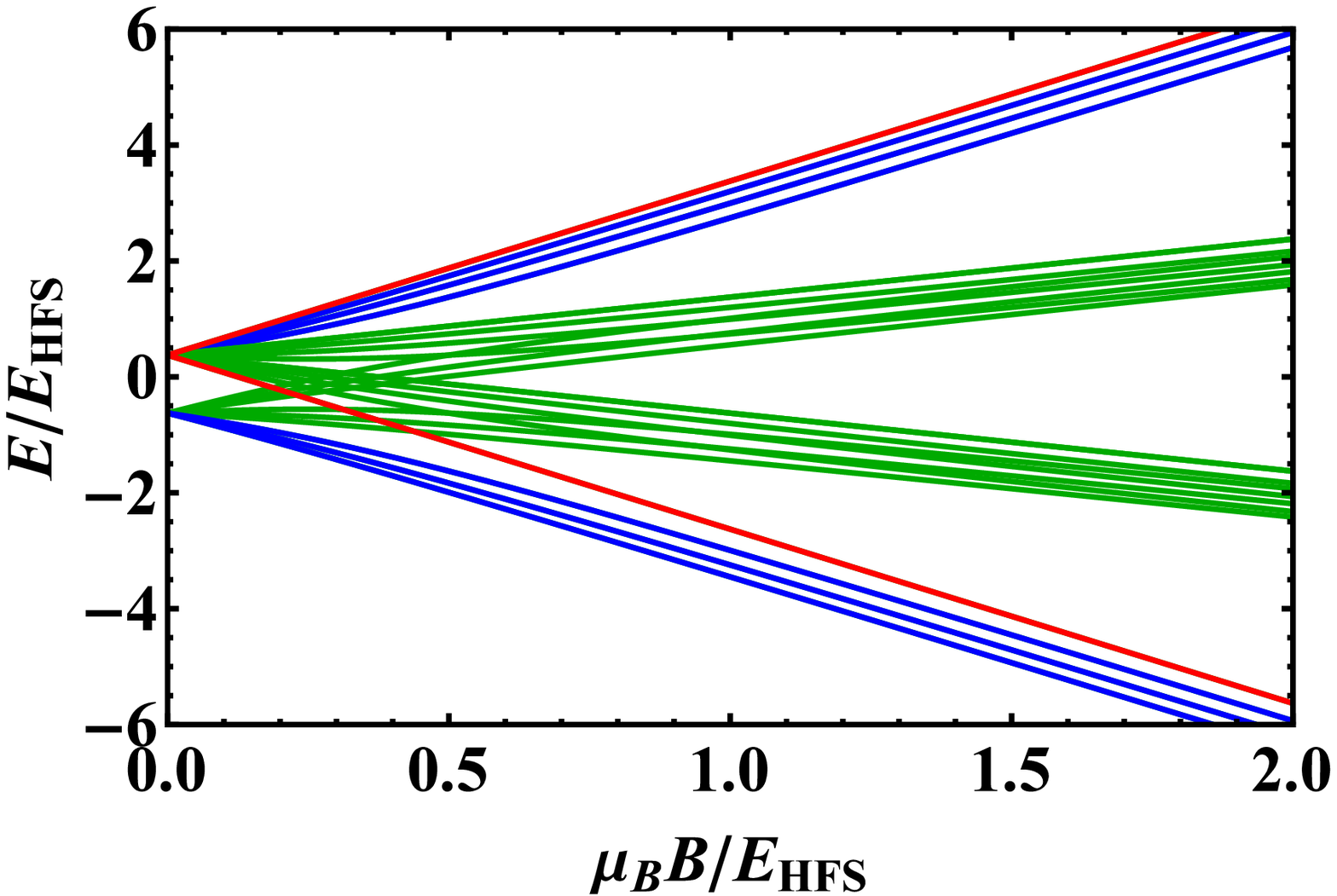}
\caption{(Color online) The energy spectrum of $^4$He$^*$+$^{87}$Rb. The blue lines indicate spin-state combinations for which $\varphi_D=0$ in the limit of high magnetic fields, the green lines $\varphi_D=1/3$ or 2/3. The red lines are the two doubly spin-stretched state combinations, for which $\varphi_D=0$ for each magnetic field.
\label{HeRbstates}}
\end{figure}

The energy spectrum of $^4$He$^*$+$^{87}$Rb is shown in Fig.~\ref{HeRbstates}, which contains 24 atom pair states. The energy and magnetic field are rescaled with the hyperfine energy, such that the spectrum applies to all $i=3/2$ alkali atoms (apart from a very small correction due to the nuclear g-factor). At high magnetic fields one recognizes four groups of states, corresponding to $m_S=-3/2,-1/2,1/2,3/2$. The outer two groups with $|m_S|=3/2$ are purely quartet, while the inner two groups $|m_S|=1/2$ have both doublet and quartet character. The blue lines indicate spin-state combinations for which $\varphi_D\rightarrow0$ in the limit of high magnetic fields, the green lines $\varphi_D=1/3$ or 2/3. At low magnetic fields these spin-state combinations have $\varphi_D=q/6$, where $q=0,\ldots, 4$. The red lines are the two doubly spin-stretched state combinations, for which $\varphi_D=0$ for any magnetic field. 

\begin{figure}[h]
\includegraphics[width=17cm]{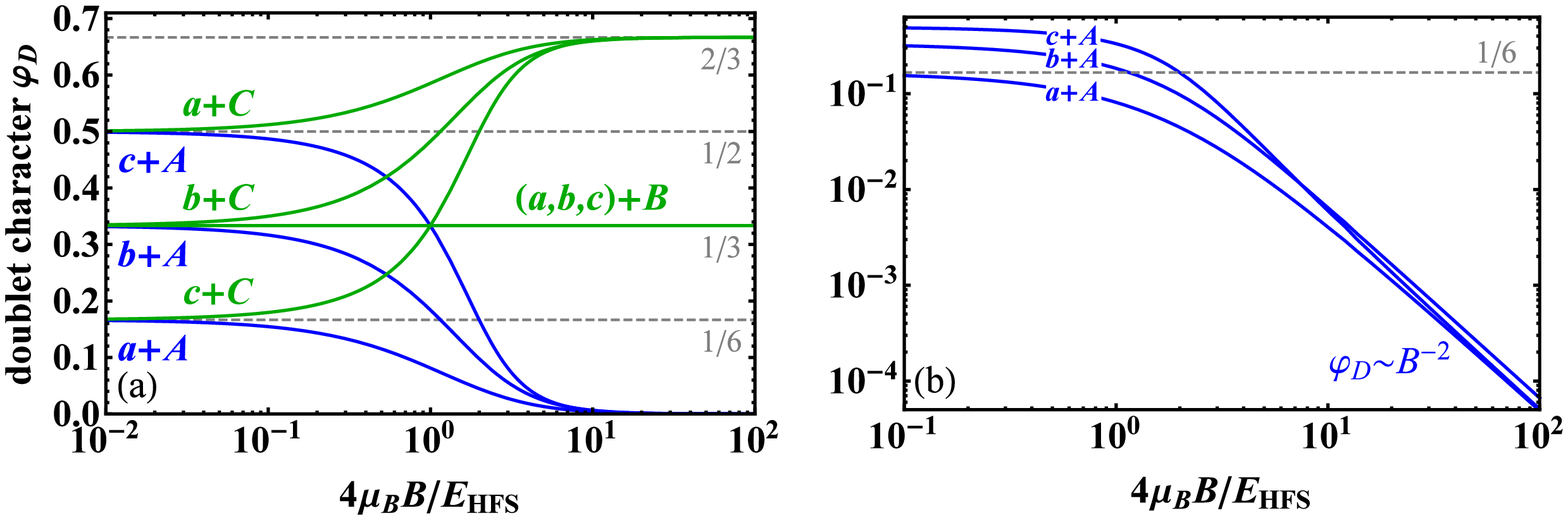}
\caption{(Color online) (a) Doublet character $\varphi_D$ as function of magnetic field for all spin-state combinations involving the $f=1$ hyperfine state of an $i=3/2$ alkali atom (labeling as shown in Fig.~\ref{Zeeman}); (b) same as (a) but showing $\varphi_D$ for $(a,b,c)+A$ on a logarithmic scale.
\label{PhiDB}}
\end{figure}

In Fig.~\ref{PhiDB}(a) $\varphi_D$ is plotted as function of magnetic field for all spin-state combinations involving the $f=1$ hyperfine state of an $i=3/2$ alkali atom, showing a smooth transition between the low and high magnetic field limits. In Fig.~\ref{PhiDB}(b) $\varphi_D$ is plotted on a logarithmic scale for $(a,b,c)+A$, emphasizing the asymptotic $\varphi_D \sim B^{-2}$ behavior for $4\mu_B B/E_{\rm HFS}>>1$.

\section{Finite temperature corrections}

\begin{figure}
\includegraphics[width=17cm]{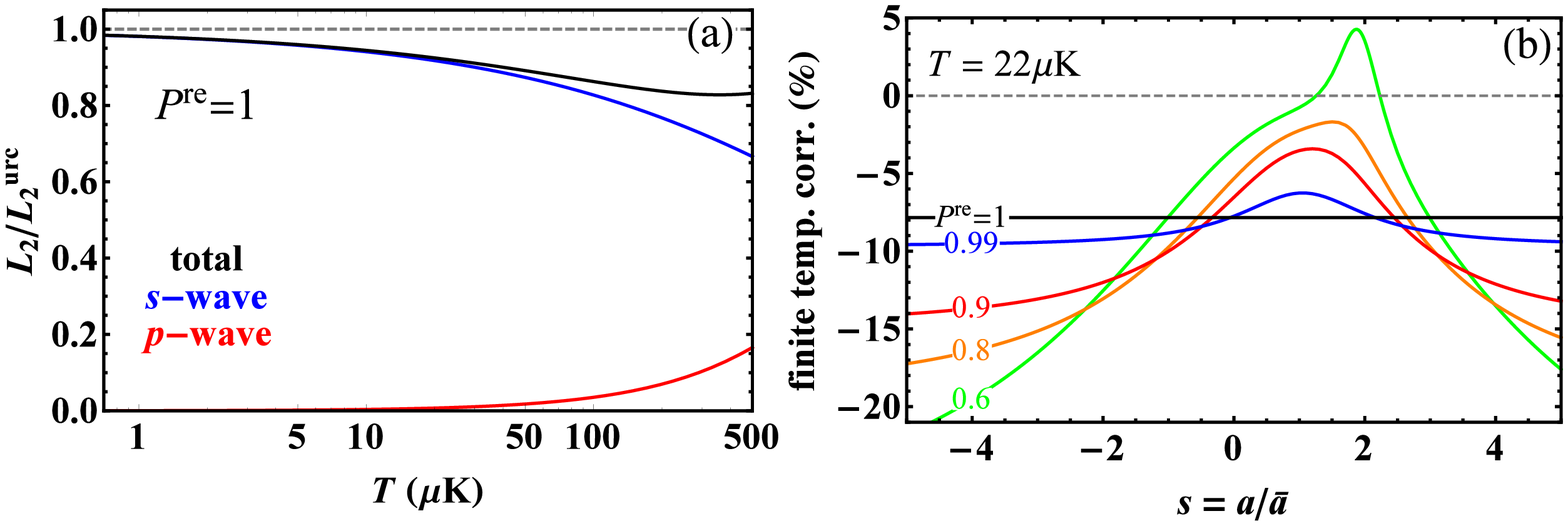}
\caption{(Color online) (a) Universal loss rate as function of temperature \cite{gao2010umf}, indicating the $s$- and $p$-wave contribution. (b) Finite temperature corrections to $L_2$ at 22~$\mu$K for the non-universal case \cite{jachymski2013qto,jachymski2014qdm} with respect to the zero-temperature limit (Eq.~\ref{nurc}).
\label{Edep}}
\end{figure}

The universal and non-universal MQDT loss rates given in Eq.~\ref{urc} and Eq.~\ref{nurc} are the zero-temperature limits. Even though our effective collision temperature of $T=\mu (T_{\rm He^*}/m_{\rm He^*}+T_{\rm Rb}/m_{\rm Rb})\approx T_{\rm He^*}=22$~$\mu$K is far below the van der Waals energy and $p$-wave centrifugal barrier ($k_B\times3.1$~mK and $k_B\times3.4$~mK, respectively), it may be necessary to include a finite temperature correction. For the universal case, $P^{re}=1$, the energy dependent, thermal averaged, loss rate is given in Ref.~\cite{gao2010umf} and is shown in Fig.~\ref{Edep}(a), specifying the contributions from the $s$-wave and $p$-wave scattering. The loss rate is reduced by 8\% at 22~$\mu$K compared to the zero-temperature limit. 

For the non-universal case, $P^{re}<1$, the energy dependence itself also depends on $P^{re}$ and $a$ \cite{jachymski2013qto,jachymski2014qdm}, most notably in case of a potential resonance ($|a|>>\bar{a}$) or shape resonances \cite{jachymski2014qdm}, such as $p$-wave shape resonances, correlated with $a \approx 2\bar{a}$. The finite temperature corrections to Eq.~\ref{nurc} at 22~$\mu$K as function of $s=a/\bar{a}$ for several values of $P^{re}$ are shown in Fig.~\ref{Edep}(b). 

\section{Spin-exchange collisions}

For some spin-state combinations exothermic spin-exchange collisions (i.\,e.\,Zeeman and hyperfine state changing collisions) are spin-allowed and could in principle have a significant contribution to the total two-body loss rate. A simple expression of the spin-exchange rate coefficient in the zero-temperature limit has been derived for the case of atomic hydrogen (and also directly applicable to alkali atoms), namely Eq.~38 of Ref.~\cite{stoof1988sea}. For the case of He$^*$+alkali collisions this expression becomes:
\begin{equation}
L_2^{\rm ex}=\pi \sqrt{\frac{2\Delta E}{\mu}} \left(a_Q-a_D\right)^2 \left|\left\langle \psi_f|P_Q-P_D|\psi_i\right\rangle \right|^2,
\end{equation}
where $\Delta E$ is the energy gain, $\mu$ is the reduced mass, $a_Q$ ($a_D$) is the quartet (doublet) scattering length, $\psi_i$ ($\psi_f$) is the initial (final) spin-state, and $P_Q$ ($P_D$) the projection operator on the quartet (doublet) spin subspace. 

\begin{figure}
\includegraphics[width=8.5cm]{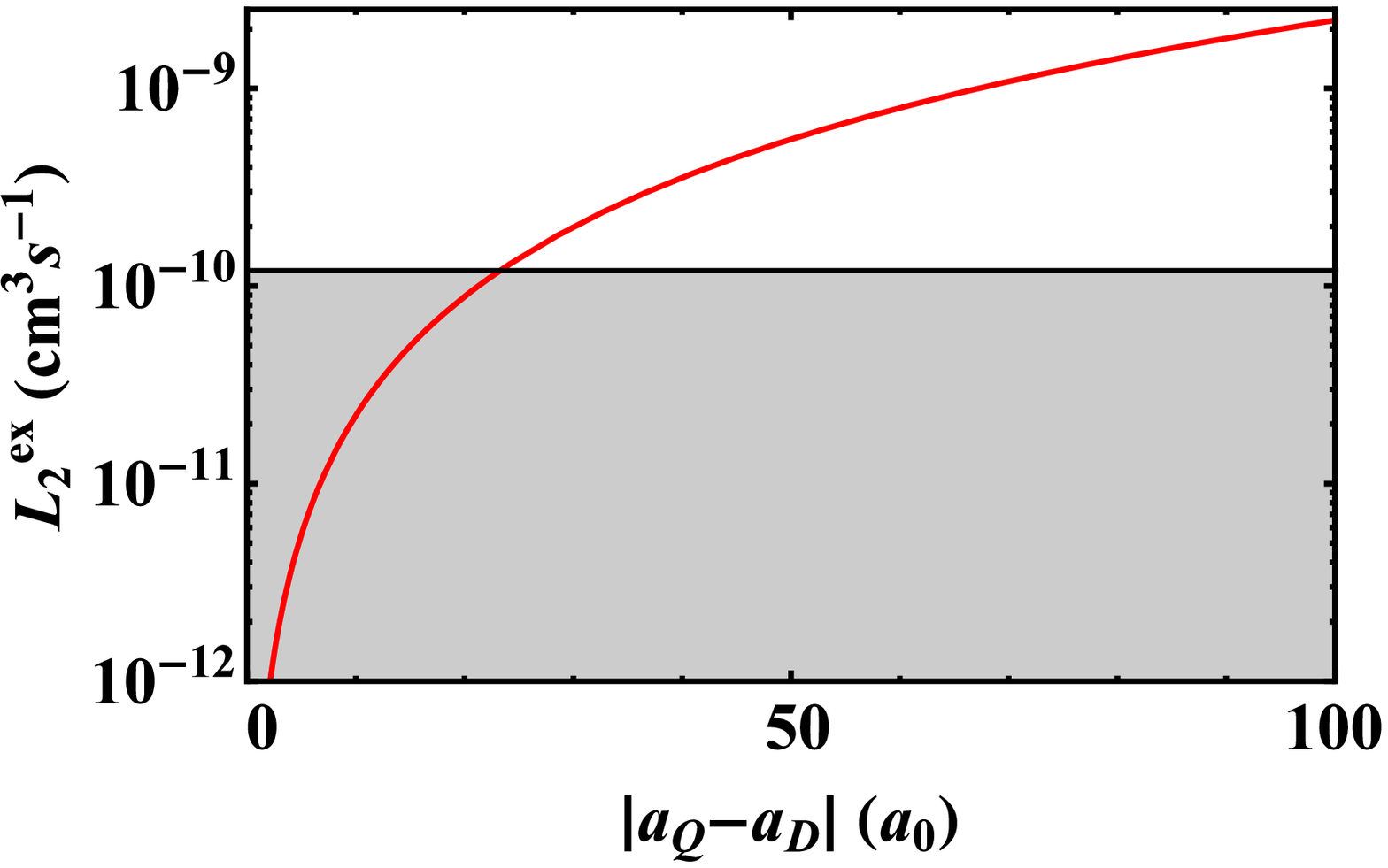}
\caption{(Color online) Loss rate $L_2^{rm ex}$ due to hyperfine changing collisions for the $h+A$ mixture as function of the difference between the doublet and quartet scattering lengths $|a_Q-a_D|$. The gray area indicates a possible contribution to the measured total two-body loss.
\label{L2HCC}}
\end{figure}

For low magnetic fields the spin-factor $\left|\left\langle \psi_f|P_Q-P_D|\psi_i\right\rangle \right|^2$ can be expressed in terms of Clebsch-Gordan coefficients. For the $^{4}$He$^*$+$^{87}$Rb case the spin-state is given by $|\psi\rangle=|s_1, m_{s_1}; f_2, m_{f_2}\rangle$. Defining
\begin{equation}\label{zeta}
\zeta_\alpha(S, m_S)\equiv\sum_{m_{s_2}}\sum_{m_{i_2}}{C_{s_2, m_{s_2}; i_2, m_{i_2}}^{f_2, m_{f_2}}C_{s_1, m_{s_1}; s_2, m_{s_2}}^{S, m_S}},
\end{equation}
where $\alpha=\{s_1, m_{s_1}; f_2, m_{f_2}; s_2, i_2\}$ is the set of spin quantum numbers of state $\psi$, then 
\begin{equation}\label{proj}
\left\langle \psi_f|P_S|\psi_i\right\rangle=\sum_{m_{S}}\zeta_i(S, m_S)\zeta_f(S, m_S).
\end{equation}
For the $h+A$ spin-state combination, in which the spin-allowed spin-exchange process is the hyperfine changing collision process $h+A\rightarrow a+B$, $\left|\left\langle \psi_f|P_Q-P_D|\psi_i\right\rangle \right|^2=2/3$. In Fig.~\ref{L2HCC} $L_2^{\rm ex}$ for $h+A$ is shown as function of $|a_Q-a_D|$, which together with our derived upper limit of $1.2\times 10^{10}$~cm$^3$s$^{-1}$ gives the constraint $|a_Q-a_D|< 23 a_0$ to the difference between the doublet and quartet scattering length.

At finite temperature the spin-exchange rate is given by Eq.~36 of Ref.~\cite{stoof1988sea}, involving the (energy-dependent) phase shifts instead of the scattering lengths. Finite temperature corrections to the zero-temperature limit are expected to be small in our case, as our temperature of 22~$\mu$K is well below the van der Waals energy and $p$-wave centrifugal barrier. The only exceptions would be in case of a potential resonance ($|a|>>\bar{a}$) or a $p$-wave shape resonance ($a \approx 2\bar{a}$) in one of the two potentials, which is not the case for the quartet potential \cite{knoop2014umo,kedziera2015aii}. The knowledge of the doublet potential is very limited, and the doublet scattering length is unknown. However, the relatively small total two-body loss rate for $h+A$, in combination with our analysis of the PI loss rate, does not support the presence of a potential resonance or $p$-wave shape resonance. 

\section{Overview of loss rates}

An overview of the measured and derived two-body loss rates are given in Table~\ref{table}. We have also added the two-body unitarity limit, 
\begin{equation}\label{unitarity}
L_2^{\rm unitarity}(T)=\frac{\sqrt{2\pi}\hbar^2}{\mu^{3/2}\sqrt{k_B T}}\approx\frac{1.5\times 10^{-8}}{\sqrt{T(\mu {\rm K})}}{\rm cm}^3{\rm s}^{-1}
\end{equation}
which at our temperature is more than one order of magnitude larger than our measured loss rates. Compared to other systems, here the unitarity limit is relatively high due to the small reduced mass $\mu$.

\begin{table}[h]
\caption{Overview of the measured and derived two-body loss rates for $^4$He$^*$+$^{87}$Rb, and comparison with the universal loss rate constant and unitarity limit (all units of $10^{-10}$~cm$^3$s$^{-1}$).}\label{table}
\begin{ruledtabular}
\begin{tabular}{cccc|c|cc|c}
$L_2^{h+C}$	&	$L_2^{a+A}$	&	$L_2^{a+C}$	&	$L_2^{h+A}$	&	$L_2^{\rm PI}$ & 	\multicolumn{2}{c|}{$L_2^{\rm urc}$} & $L_2^{\rm unitarity}$\\
 & & & &  & $T=0$ & $T=22$~$\mu$K & $T=22$~$\mu$K \\
\hline
 & & & & & & & \\
	$<0.013$	&	$0.53^{+0.20}_{-0.17}$	& $2.3^{+1.2}_{-0.9}$	& $2.2^{+0.8}_{-0.7}$	& $3.4^{+0.8}_{-0.7}$	 & 4.5 & 4.2	& 32 \\  
	 & & & & & & & \\
\end{tabular}
\end{ruledtabular}
\end{table}

\end{document}